\documentclass[12pt]{article}

\usepackage[dvips]{graphicx}

\def\dspace{\baselineskip = 0.30in}

\def\lapproxeq{\lower .7ex\hbox{$\;\stackrel{\textstyle
<}{\sim}\;$}}
\def\gapproxeq{\lower .7ex\hbox{$\;\stackrel{\textstyle
>}{\sim}\;$}}

\begin{document}

\dspace

\begin{titlepage}
\begin{flushright}
%\preprint{
BA-02-25\\
%March 2002\\
%}
\end{flushright}
\vskip 2cm
\begin{center}
%%\title
{\Large\bf 
A model with bulk and brane gauge kinetic terms   
and SUSY breaking without hidden sector
}
\vskip 1cm
{\normalsize\bf
%\author{
Bumseok Kyae\footnote{bkyae@bartol.udel.edu} 
%and 
%\footnote{}
}
\vskip 0.5cm
{\it Bartol Research Institute, University of Delaware, \\Newark,
DE~~19716,~~USA\\[0.1truecm]}

%
%%\maketitle

\end{center}
\vskip .5cm

%\date{\today}
%\pacs{PACS: 11.25.Mj, 12.10.Dm, 98.80.Cq}

\begin{abstract}
We derive various solutions for the gauge
field and the gaugino when there are 
both 5D bulk kinetic terms and 4D brane kinetic terms.  
% on an $S^1/Z_2$ fixed point.  
%
%4D brane kinetic terms with the gauge coupling 
%$e^2$ on an $S^1/Z_2$ fixed point
%as well as 5D bulk kinetic term with $g^2$.        
Below the compactification scale $1/y_c$, 
gauge interaction by the massless mode
is universal, independent of the locations of sources, 
but above $1/y_c$, the interaction distinguishes their locations.  
We consider the $S^1/(Z_2\times Z_2')$ orbifold compactification, 
in which $N=2$ SUSY and $SU(2)$ gauge symmetry 
break down to $N=1$ and $U(1)$, respectively.  
While the odd parity gauge fields under $Z_2$, 
$A^1_\mu$ and $A^2_\mu$, interact with the bulk gauge coupling $g^2$, 
$A^3_\mu$ at low energy interacts with  
$e^2g^2/(e^2+g^2)$ due to the brane kinetic term with the coupling $e^2$.  
Even if $g^2$, which is asymptotically free, blew up at the scale  
$1/y_c<\mu < \Lambda_{\rm cutoff}$, 
$e^2g^2/(e^2+g^2)$ could remain small at low energy 
by brane matter fields' contribution.  
The condensations of the gauginos $\Psi^1$ and $\Psi^2$ generate  
soft mass term of $\Psi^3$ by gravity mediation.   
 
\end{abstract}
\end{titlepage}

\newpage

%%%%%%%%%
%%%%%%%%%

%%%%%%%%%%%%%%%%%%%%%%%%%%%%%%%%%%%%%%%%%%%%%%%%%%%%%%%%%%%%%%%%%%%%%%%%%%%
\section{Introduction}

For the last seven years, the brane world idea 
has been one of the major research 
fronts in particle physics, because the idea    
%One of the reasons is that the idea 
%
%that  
%some fields reside only on the sub-space-time (or brane) embedded in higher 
%dimensional space-time ($D>4$) whereas other fields propagate 
%in the whole space-time (bulk), 
%
provides possibilities to resolve 
many theoretically difficult problems in particle physics 
in geometrical ways.   
In the earlier stage of the brane revolution, it was discussed that  
the brane scenario could be applied to resolve  
various scale problems  
like the gauge hierarchy problem \cite{add,rs} and 
the $\mu$ problem \cite{mu}.  
In the setup where all kinds of the standard model fields live on a 4D brane 
while gravity can propagate in the whole higher dimesional space-time (bulk),  
the fundamental scale  
%which was belived to be in around the Planck scale, 
may be lowered to the grand unification scale 
or even to TeV scale by assuming large volume of the extra dimensional space.  
Such a setup was utilized to resolve  
%scale problems in particle physics like 
the gauge hierarchy problem \cite{add} and 
the string-GUT scale unification problem \cite{horava}.  
%
%In the setup where supersymmetry breaking sectors are localized on branes, 
%SUSY breaking scales could be non-universal at Planck scale 
%even within grvity mediation SUSY breaking scenario \cite{kyae}.   
%

Recently, it was pointed that the orbifold compactification gives 
an efficient mechanism for breaking supersymmetry (SUSY) and/or gauge symmetry 
in the context of the higher dimensional supersymmetric grand unified theory 
(GUT) \cite{kawamura}.  
In orbifold compactifications, a presumed discrete symmetry 
of the extra dimensional space is imposed to the field space.    
Then, only invariant bulk fields under the symmetry survive at low energy, 
and can couple to brane fields on lower dimensional 
orbifold fixed points (brane).  
In $S^1/(Z_2\times Z_2')$ orbifold compactification models, 
the original gauge group ${\cal G}$ that the bulk Lagrangian respects 
breaks down to a sub-group ${\cal H}$ commuting with $Z_2$, 
and only one half of the supersymmetries introduced in the bulk action 
survive at low energy because of the other $Z_2'$.     
Then, the required gauge symmetry on the branes  
is ${\cal H}$ rather than ${\cal G}$ \cite{hebecker}.  
A resolution to the doublet-triplet splitting problem 
in GUT employing this mechanism was proposed in \cite{kawamura}.   

In this paper we will study extensively the model  
in which the kinetic terms of the gauge field(s) and 
gaugino(s) with even parity are present both in the bulk and  
on the brane(s) ($Z_2$ orbifold fixed point(s)), 
in the $S^1/(Z_2\times Z_2')$ orbifold compactification \cite{dvali,etc}.  
Only if the brane kinetic term respects $N=1$ SUSY and 
the gauge symmetry ${\cal H}$, are such terms harmless.      
Moreover, brane kinetic terms, generally, 
could be generated radiatively on the brane, 
even if it is absent in the tree level action \cite{dvali}.  
In this paper, however, we will assume 
the brane kinetic term is not small from the beginning.  
In this setup, we will derive the gauge field and 
the gaugino propagators, and 
discuss a SUSY breaking scenario with the propagators.  

\section{Gauge field solutions}

In the presence of the kinetic terms of the gauge field and gaugino 
on the brane at $y=0$ as well as in the bulk, the action is 
\begin{eqnarray} \label{action}
S&=&\int d^4x\int^{y_c}_{-y_c}dy~\bigg[
\frac{1}{2y_c}~\frac{-1}{4g^2}F_{MN}F^{MN}
+\delta(y)\frac{-1}{4e^2}F_{\mu\nu}F^{\mu\nu}-\bigg(A^M J_M\bigg) 
\nonumber \\ 
&&~~~+\frac{1}{2y_c}~\frac{i}{g^2}\overline{\Psi}D_5\hspace {-4mm}/~\Psi
+\delta(y)\frac{i}{e^2}\overline{\Psi}D_4\hspace {-4mm}/~\Psi 
-\bigg(\overline{\Psi}\eta+\overline{\eta}\Psi\bigg)\bigg] ~,  
\end{eqnarray}
in which $J_M$ and $\eta$ are vector-like and fermionic sources that are 
coupled to the gauge boson and gaugino, respectively.  
$g^2$ and $e^2$ are the couplings of the bulk and brane  gauge kinetic terms.  
In this paper, $M$, $N$ indicate 5D space-time, $(x^\mu,x^5=y)$, where 
$\mu=0,1,2,3$.   
When the source is localized on the brane, 
$J_M(x,y)=\delta_M^\mu J_\mu(x)\delta(y)$,  
the equation of motion of the gauge field,   
derived from the action (\ref{action}) is \cite{dvali}
\begin{eqnarray}\label{bosoneq}
&&\bigg[k^2+\partial^2_y
+\delta(y)\frac{2y_cg^2}{e^2}~k^2\bigg]\tilde{A}_\mu(k,y) 
%-ik_\mu\partial_y\tilde{A}_5\bigg] 
=-2y_cg^2\tilde{J}_\mu (k)\delta (y) ~,  \\ 
&&\bigg[k^2+\partial^2_y\bigg]\tilde{A}_5(k,y)-i\partial_y\bigg[
\delta(y)\frac{2y_cg^2}{e^2}~k^\mu\tilde{A}_\mu(k,y)\bigg]=0  ~, 
\label{bosoneq2}
\end{eqnarray}
where we take the gauge condition, 
$\partial^MA_M/2y_cg^2+\delta(y)\partial^\mu A_\mu/e^2=0$.   
The fields with `tilde's indicate the fields in $(k,y)$ space, 
obtained by the Fourier transformation.  
In Eq.~(\ref{bosoneq}), the first and the second terms 
of the left hand side come from the bulk kinetic term  
in the action (\ref{action}), 
and the third term is from the brane kinetic term localized at $y=0$.   
In Eq.~(\ref{bosoneq}), because the right hand side is proportional 
only to $\delta (y)$, it is necessary to kill the first two terms
in the left hand side.  
Hence, the solution should be a combination of the trigonometric functions,  
${\rm sin}ky$ and ${\rm cos}ky$.    
The single-valued condition of the solution on $S^1$ 
requires different combinations of ${\rm sin}ky$ and ${\rm cos}ky$ 
at the two intervals $(-y_c,0)$ and $(0,y_c)$, which potentially gives 
rise to singular terms proportional to $\delta (y)$ and $\delta(y-y_c)$.    
A unique combination that does not generate any 
singular term at $y=y_c$ is ${\rm cos}k|y-y_c|$, and the solution 
is given by     
\begin{eqnarray} \label{1stsol1}
\tilde{A}^{b1}_\mu(k,y)=-f(k)~{\rm cos}k|y-y_c|\cdot 2y_cg^2\tilde{J}_\mu(k) 
~, 
\end{eqnarray}
where our definition of ``$|y-y_c|$'' is shown in Fig. 1-(a), and 
$f(k)$ is determined to satisfy the equation of motion,   
\begin{eqnarray} \label{1stsol2}
f(k)=\frac{1}
{2k~{\rm sin}ky_c+\frac{2g^2}{e^2}k^2y_c~{\rm cos}ky_c} ~.   
\end{eqnarray}
\begin{figure}
%[t]
\begin{center}
\includegraphics[width=120mm]{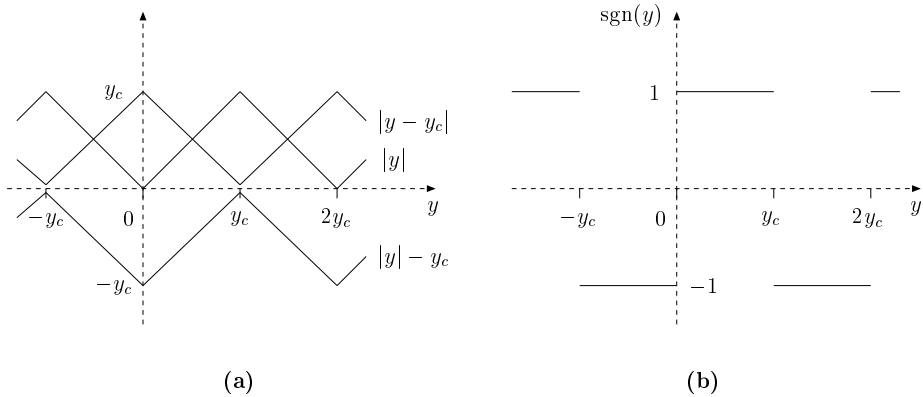}
\end{center}
\caption{The definitions of the functions, 
${\bf (a)}$ $|y-y_c|$, $|y|$, and $|y|-y_c$, 
and ${\bf (b)}$ ${\rm sgn}(y)$ in this paper.  They all are periodic in the 
compact extra dimension.  }
\end{figure}

With the solution Eq. (\ref{1stsol1}), it can be also proved that 
$\tilde{A}_5(k,y)$ in Eq. (\ref{bosoneq2}) should be an odd function 
proportional to ${\rm sgn}(y)k^\mu\tilde{A}_\mu$, 
where our definition of ``${\rm sgn}(y)$'' is shown in Fig. 1-(b).  
    
The expression of ${\rm cos}k|y-y_c|$ in Eq.~(\ref{1stsol1}) 
contains Kaluza-Klein (KK) modes' contributions.  
To compare the solution with the ordinary solution in the absence of 
the brane kinetic term, it is necessary 
to expand ${\rm cos}k|y-y_c|$ in the fifth momentum space 
with the bases of $e^{ik_5y}$($=e^{in\pi y/y_c}$),  
\begin{eqnarray}
{\rm cos}k|y-y_c|=\frac{1}{2y_c}\sum_{n=-\infty}^{\infty}\tilde{g}_n
e^{in\pi y/y_c} ~, 
\end{eqnarray}
where $\tilde{g}_n$ is given by 
\begin{eqnarray}
\tilde{g}_n=\int_{-y_c}^{y_c}dy e^{-in\pi y/y_c}{\rm cos}k|y-y_c|=
\frac{2k~{\rm sin}ky_c}{k^2-(n\pi/y_c)^2} ~.  
\end{eqnarray}
Hence, the solution in $(x,y)$ space is 
\begin{eqnarray} \label{xygauge}
A^{b1}_\mu(x,y)=-\int \frac{d^4k}{(2\pi)^4}\sum_{k_5} 
\frac{e^{ikx}e^{ik_5y}}{k^2-k_5^2}~\frac{e^2g^2 ~\tilde{J}_\mu (k)}
{e^2+g^2ky_c~{\rm cot}ky_c} ~.  
\end{eqnarray}
We note that when both the bulk and  the brane kinetic terms  
are present and a source is localized on the brane, 
the bulk gauge coupling $g^2$ is effectively replaced 
by $e^2g^2/(e^2+g^2ky_c{\rm cot}ky_c)$.  
In the limit $e^2\longrightarrow \infty$, which corresponds to 
the case in which the brane kinetic term is absent 
in the action (\ref{action}), 
the solution reduces to the ordinary solution.    
In the low energy limit $ky_c\longrightarrow 0$ 
(or $ky_c{\rm cot}ky_c\longrightarrow 1$), 
the effective coupling is given by $e^2g^2/(e^2+g^2)$.  It is also 
consistently checked in the action after integrating out its $y$ dependence.  
In the Euclidian space, $ky_c{\rm cot}ky_c$ becomes $k_E{\rm coth}k_Ey_c$.  
Thus, the gauge interaction at high energy is more suppressed 
by the factor $e^2/(g^2k_Ey_c)$ 
compared to that expected in the ordinary theory 
without the brane kinetic term \cite{dvali}.   

Similarly, we can directly derive the solution for the case 
when a brane kinetic term and a source term are located at $y=y_c$,   
\begin{eqnarray}\label{soly=c}
\tilde{A}^{b2}_\mu(k,y)=-\frac{{\rm cos}k|y|\cdot 2y_cg^2\tilde{J}_\mu(k)}
{2k~{\rm sin}ky_c+\frac{2g^2}{e_c^2}k^2y_c~{\rm cos}ky_c} ~,
\end{eqnarray}
where $e_c$ is the gauge coupling of a gauge kinetic term localized 
at $y=y_c$.  
Our definition of ``$|y|$'' is also shown in Fig.1-(a). 
The solution (\ref{soly=c}) does not generate a singular term proportional 
to $\delta (y)$.  
The function ${\rm cos}k|y|$ expands in the fifth momentum space 
in the following way:   
\begin{eqnarray}
{\rm cos}k|y|=\frac{1}{2y_c}\sum_{n=-\infty}^{\infty}
e^{in\pi (y-y_c)/y_c}\frac{2k~{\rm sin}ky_c}{k^2-(n\pi/y_c)^2} ~.  
\end{eqnarray}
We note again that for $e_c^2>>1$ and $1/y_c>>k$, the solution reduces to 
the ordinary one with the coupling $g^2$.  

Let us consider the case containing two localized sources 
both at $y=0$ and at $y=y_c$, and a brane gauge kinetic term at $y=0$,   
\begin{eqnarray}\label{both}
\bigg[k^2+\partial^2_y+\delta(y)\frac{2y_cg^2}{e^2}k^2\bigg]
\tilde{A}_\mu(k,y)
=-2y_cg^2\bigg(\tilde{J}_\mu^{b1}(k)\delta (y)
+\tilde{J}_\mu^{b2} (k)\delta (y-y_c)\bigg)
\end{eqnarray}
As a trial solution, we take $\tilde{A}_\mu(k,y)=\tilde{A}^{b1}_\mu(k,y)+
\tilde{A}^{b2}_\mu(k,y)$ with $e_c\rightarrow \infty$, 
where $\tilde{A}^{b1}_\mu(k,y)$ and $\tilde{A}^{b2}_\mu(k,y)$ are given 
from Eq.~(\ref{1stsol1}) and (\ref{soly=c}), respectively. 
Then the left hand side of Eq.~(\ref{both}) gives
\begin{eqnarray}
-2y_cg^2\bigg(\tilde{J}_\mu^{b1}(k)\delta(y)
+\tilde{J}_\mu^{b2}(k)\delta(y-y_c)\bigg) ~ 
-~2y_cg^2\delta(y)~\frac{g^2}{e^2}~\tilde{I}^{b2}_\mu(k) ~, 
\end{eqnarray}
where $\tilde{I}^{b2}_\mu(k)$ is defined as   
\begin{eqnarray}
\tilde{I}^{b2}_\mu(k)&\equiv&\sum_{n=-\infty}^{\infty}
\frac{(-1)^nk^2}{k^2-(n\pi/y_c)^2}~\tilde{J}^{b2}_\mu (k)  ~.  
\end{eqnarray} 
The effective 4D theory at low energy could be obtained by decoupling or 
truncating all of heavy KK modes. 
Then, $\tilde{I}^{b2}_\mu(k)$ is given just by $\tilde{J}^{b2}_\mu(k)$.    
To cancel the unwanted third term, the source term at $y=0$ should 
be shifted to $\tilde{J}^{b1}_\mu(k)-(g^2/e^2)~\tilde{I}^{b2}_\mu$.  
Thus, the solution is  
\begin{eqnarray}
A_\mu(x,y)&=&A_\mu^{b1}(x,y)+A_\mu^{b2}(x,y) ~, \\
A_\mu^{b1}(x,y)&=&-\int \frac{d^4k}{(2\pi)^4}\sum_{n}
\frac{e^{ikx}e^{in\pi y/y_c}}{k^2-(n\pi/y_c)^2}
~\frac{(e^2g^2 \tilde{J}_\mu^b (k)-g^4
\tilde{I}_\mu^{b2}(k))}{e^2+g^2ky_c~{\rm cot}ky_c} ~, \\
A_\mu^{b2}(x,y)&=&-\int\frac{d^4k}{(2\pi)^4}\sum_{n}
\frac{e^{ikx}e^{in\pi(y-y_c)/y_c}}{k^2-(n\pi/y_c)^2}~ 
g^2\tilde{J}_\mu^{b2} (k) ~.  
\end{eqnarray}
We note that the two point Green functions of  
the gauge interaction between two brane sources at $y=0$ are different from 
those between those at $y=y_c$.  
However, if $ky_c<<1$, at low energy where massive KK particles are 
decoupled, their effective gauge couplings $(e^2g^2/(e^2+g^2))$ 
become universal and the propagator reduces to ordinary 4D propagator form.   

Similarly, we can also easily derive the solution 
in the presence of a bulk source 
as well as a source localized at $y=0$,  
%
%\begin{eqnarray}
%\bigg[k^2+\partial^2_y+\delta(y)\frac{2y_cg^2}{e^2}k^2\bigg]
%\tilde{A}_\mu(k,y)
%=-2y_cg^2\bigg(\tilde{J}_\mu^B(k,y)+\tilde{J}_\mu^b (k)\delta (y)\bigg)
%\end{eqnarray}
%
\begin{eqnarray} \label{bosonb}
A_\mu(x,y)&=&A_\mu^B(x,y)+A_\mu^b(x,y) ~, \\  \label{bosonm}
A_\mu^B(x,y)&=&-\int\frac{d^4k}{(2\pi)^4}\sum_{k_5}
\frac{e^{ikx}e^{ik_5y}}{k^2-k_5^2}\cdot g^2\hat{J}_\mu^B(k,k_5) ~, \\
A_\mu^b(x,y)&=&-\int \frac{d^4k}{(2\pi)^4}\sum_{k_5}
\frac{e^{ikx}e^{ik_5y}}{k^2-k_5^2}~\frac{(e^2g^2 \tilde{J}_\mu^b (k)-g^4
\tilde{I}_\mu^B(k))}{e^2+g^2ky_c~{\rm cot}ky_c} ~,  
\end{eqnarray}
where $\hat{J}_\mu^B(k,k_5)$ and $\tilde{I}_\mu^B(k)$ are defined as 
\begin{eqnarray}
\tilde{J}_\mu^B(k,y)&\equiv& \frac{1}{2y_c}\sum_{k_5} \hat{J}_\mu^B(k,k_5)
e^{ik_5y} ~, \\
\tilde{I}_\mu^B(k)&\equiv& \sum_{k_5} \frac{k^2}{k^2-k_5^2} ~ 
\hat{J}_\mu^B(k,k_5) ~.  \label{bosone}
\end{eqnarray}
Thus, the propagator mediating two bulk sources and 
two localized sources in 4D space-time are, respectively,  
\begin{eqnarray}\label{green}
&&\sum_{n}\frac{-ig_{\mu\nu}g^2}{k^2-(n\pi/y_c)^2}\bigg(1
-\frac{g^2}{e^2+g^2ky_c~{\rm cot}ky_c}
\sum_{n'}\frac{k^2}{k^2-(n'\pi/y_c)^2}\bigg)  ~, \\
&&\sum_{n}\frac{-ig_{\mu\nu}}{k^2-(n\pi/y_c)^2}~\frac{e^2g^2}{e^2+g^2ky_c~
{\rm cot}ky_c} ~. \label{green2} 
\end{eqnarray}
We note again that if all KK modes are decoupled,
in the limit $ky_c\rightarrow 0$, they become the same as each other,
\begin{eqnarray}
\frac{e^2g^2}{e^2+g^2}~\frac{-ig_{\mu\nu}}{k^2} ~,
\end{eqnarray}
where $e^2g^2/(e^2+g^2)$ is the effective coupling at low energy.  
Thus, {\it even if $g^2$ is extremely large, 
the effective gauge couplings of the two propagators could be small 
if $e^2$ is tiny. }  

When two branes at $y=0$ and $y=y_c$ contain both a brane kinetic term 
and a source term, respectively, it is difficult to obtain the exact    
solution.  However, if $g^2<<e_1^2, e_2^2$ or $g^2,e_1^2<<e_2^2$, we can 
get approximate solutions around the solutions when   
$e_1^2, e_2^2\rightarrow \infty$ or $e_2^2\rightarrow \infty$ 
which are provided already.  
We can check that at low energy the approximate solutions also 
maintain the universality of the gauge couplings.  

Now let us discuss for a while the property of the function,  
$g^2/(e^2+g^2ky_c~{\rm cot}ky_c)$ appearing in Eq.~(\ref{green}) and 
(\ref{green2}).   
The function $g^2/(e^2+g^2ky_c~{\rm cot}ky_c)$ 
contains infinite number of poles, and   
it could be expanded with functions singular at each poles.    
It means that infinite number of additional massive particles 
are concerned in the propagators.     
%
%When we derive the effective theory at a scale $\mu$, 
%heavier KK modes than $\mu$ could be truncated, while lighter 
%KK modes than it would be regarded as being massless particles 
%approximately.   
%
%it is necessary to expand the fuction $g^2/(e^2+g^2ky_c~{\rm cot}ky_c)$ 
%with functions carrying only one pole among the poles of it.  
%
%
To see it clearly, let us consider the simpler case of $e^2=0$.  
It corresponds to the case that the effect of the brane kinetic term 
is extremely large.  The function $1/ky_c{\rm cot}ky_c$ is    
\begin{eqnarray}\label{1/xcotx}
\frac{1}{ky_c~{\rm cot}ky_c}=\frac{2/y_c^2}{(\pi/2y_c)^2-k^2}
+\frac{2/y_c^2}{(3\pi/2y_c)^2-k^2}+\frac{2/y_c^2}{(5\pi/2y_c)^2-k^2}+\cdots ~, 
\end{eqnarray}
where $\pi/2y_c$, $3\pi/2y_c$, $5\pi/3y_c$, and so forth are 
singular points of $1/ky_c{\rm cot}ky_c$.  
Thus, at low energy $k<<1/y_c$, it gives 
$(2/y_c^2)(2y_c/\pi)^2(1+1/3^2+1/5^2+\cdots)=(8/\pi^2)(3\pi^2B_1/4)=1$, where 
$B_1=1/6$ is a Bernoulli number.  

When we turn on $e^2$, the expansion in Eq.~(\ref{1/xcotx}) 
should be changed. 
$g^2/(e^2+g^2ky_c{\rm cot}ky_c)$ expands with the functions, 
\begin{eqnarray}
\frac{1}{ky_c{\rm cot}~ky_c}~\bigg[\bigg(\frac{e^2}{g^2}\bigg)
\frac{1}{ky_c{\rm cot}ky_c}\bigg]^n ~, 
\end{eqnarray}
in which the expansion of $1/ky_c{\rm cot}ky_c$ was given 
in Eq.~(\ref{1/xcotx}).  
Thus, the second term in the propagator Eq.~(\ref{green}) expands
\begin{eqnarray} \label{e^2}
&&\sum_{k_5,k'_5}\bigg[\frac{-ig_{\mu\nu}g^2}{k^2-k_5^2}
\frac{k^2}{k^2-k_5^{'2}}\bigg]\bigg(\frac{g^2}{e^2+g^2ky_c~{\rm cot}ky_c}
\bigg)
\nonumber \\
&&=\sum_{k_5,k'_5}\bigg[
\frac{-ig_{\mu\nu} g^2}{k^2-k_5^2}\frac{k_5^2}{k^2-k_5^{'2}}
-\frac{-ig_{\mu\nu}g^2}{k^2-k_5^{'2}}\frac{k_5^{'2}}{k^2-k_5^{'2}}\bigg]
 \\
&&~~~~~~\times \bigg(\frac{1}{ky_c~{\rm cot}ky_c}
\bigg(1-\frac{e^2}{g^2ky_c~{\rm cot}ky_c}
+\cdots \bigg)\bigg) ~.  \nonumber
\end{eqnarray}
Since $(1/ky_c{\rm cot}ky_c)^n$ can be always decompsed to a summation of 
the functions proportional to $1/[((2m+1)\pi/2y_c)^2-k^2]$ ($m=0,1,2,\cdots$),
we can conclude that even when $e^2\neq 0$ the propagator in Eq.~(\ref{green})
can be expressed always as the summation of the ordinary prapagators 
in which more heavy modes with masses $M_m=(2m+1)\pi/2y_c$ ($m=0,1,2,\cdots$)
are involved. 
Thus, at low energy $\mu <\pi/y_c$, only the mode with $k_5=k'_5=0$ 
makes a contribution to the propagator, and the part in the parenthesis 
in Eq.~(\ref{e^2}) gives the modified gauge coupling $e^2/(e^2+g^2)$.   
  
As seen in the case $e^2=0$, the case of $e^2\neq 0$ 
would also introduce more massive modes with masses $k_5'$ and $M_m$.    
However, the above results when $e^2=0$ would be still approximately valid 
at very high energy scales beyond $\pi/y_c$ even when $e^2\neq 0$.  

\section{Gaugino solutions}

For completness of our discussion, 
let us consider the fermionic cases also.   
The gaugino equation of motion in the presence of the brane kinetic term 
as well as bulk kinetric term is  
\begin{eqnarray}\label{fermeq}
\bigg[\gamma_5k\hspace {-2.3mm}/-i\partial_y
+\delta(y)\frac{2y_cg^2}{e^2}\gamma_5k\hspace {-2.3mm}/\bigg]\tilde{\Psi}(k,y)
=2y_cg^2\gamma_5\tilde{\eta}(k)\delta(y) ~, 
\end{eqnarray}
where $\gamma_5\equiv{\rm diag}\bigg(iI_{2\times 2}, -iI_{2\times 2}\bigg)$. 
For convenience, we multiplied $\gamma_5$ to the both side of the equation 
(\ref{fermeq}).   
The solution $\tilde{\Psi}(k,y)$ turns out to have the follownig form,   
\begin{eqnarray}\label{fermionsol}
\tilde{\Psi}(k,y)&=&\tilde{\Psi}^1(k,y)-\tilde{\Psi}^2(k,y) ~, \\
\tilde{\Psi}^1(k,y)&=&f(\gamma_5k\hspace{-2.3mm}/,k) ~ \label{sol1}
{\rm exp}\bigg(-i\gamma_5k\hspace {-2.3mm}/(|y|-y_c)\bigg)
\bigg[{\rm sgn}(y)+1\bigg] \cdot 2y_cg^2\gamma_5\tilde{\eta}(k) ~,\\
\tilde{\Psi}^2(k,y)&=&f(\gamma_5k\hspace{-2.3mm}/,k) ~  \label{sol2}
{\rm exp}\bigg(-i\gamma_5k\hspace{-2.3mm}/|y-y_c|\bigg)
\bigg[{\rm sgn}(y)-1\bigg] \cdot 2y_cg^2\gamma_5\tilde{\eta}(k) ~, 
\end{eqnarray}
where our definitions of ``$|y|-y_c$'' is shown in Fig. 1-(a).   
This combination also does not generate any singular term at $y=y_c$.
With $(\gamma_5 k\hspace{-2.3mm}/)^2=k^2$, 
one can find the expression of $f(\gamma_5k\hspace{-2.3mm}/,k)$ 
satisfying the equation of motion,  
\begin{eqnarray} \label{coeff2}
f(\gamma_5k\hspace{-2.3mm}/,k)=\frac{\gamma_5k\hspace{-2.3mm}/}{4k}~\frac{1}
{{\rm sin}ky_c+\frac{g^2}{e^2}ky_c~{\rm cos}ky_c} ~. 
\end{eqnarray}
%
%When we derive Eq.~(\ref{coeff}), we use the r 
%\begin{eqnarray}
%&&(\gamma_5 k\hspace{-2.3mm}/)^2=k^2 ~, \\
%&&{\rm exp}(i\gamma_5 k\hspace{-2.3mm}/y_c)+
%{\rm exp}(-i\gamma_5 k\hspace{-2.3mm}/y_c)\equiv 
%2~{\rm cos}(\gamma_5 k\hspace{-2.3mm}/y_c)= 2~{\rm cos}ky_c ~, \\
%&&{\rm exp}(i\gamma_5 k\hspace{-2.3mm}/y_c)-
%{\rm exp}(-i\gamma_5 k\hspace{-2.3mm}/y_c)\equiv 
%2i~{\rm sin}(\gamma_5 k\hspace{-2.3mm}/y_c)
%=2i\frac{\gamma_5k\hspace{-2.3mm}/}{k}{\rm sin}ky_c 
%\end{eqnarray}
%
Let us expand the $y$ dependent parts of the solution,  
Eq.~(\ref{sol1}) and (\ref{sol2}) in the fifth momemtum space,   
\begin{eqnarray}
&&{\rm exp}\bigg(-i\gamma_5k\hspace{-2.3mm}/(|y|-y_c)\bigg)
\bigg[{\rm sgn}(y)+1\bigg]
=\frac{1}{2y_c}\sum_{n}\tilde{g}_{n}^1e^{in\pi y/y_c} ~,\\
&&{\rm exp}\bigg(-i\gamma_5k\hspace{-2.3mm}/|y-y_c|\bigg)
\bigg[{\rm sgn}(y)-1\bigg]
=\frac{1}{2y_c}\sum_{n}\tilde{g}_{n}^2e^{in\pi y/y_c} ~,\\
&&\tilde{g}_{n}^1-\tilde{g}_{n}^2=4~{\rm sin}ky_c~
\frac{\gamma_5k\hspace{-2.3mm}/}{k} ~
\frac{\gamma_5 k\hspace{-2.3mm}/-n\pi/y_c}{k^2-(n\pi/y_c)^2} ~.  
\end{eqnarray}
Hence, the gaugino solution in $(x,y)$ space is  
\begin{eqnarray}
\Psi(x,y)=\int \frac{d^4k}{(2\pi)^4}\sum_{k_5}
e^{ikx}e^{ik_5y}\frac{k\hspace{-2.3mm}/-k\hspace{-2.3mm}/_5}{k^2-k_5^2}~
\frac{e^2g^2 ~\tilde{\eta}(k)}
{e^2+g^2ky_c~{\rm cot}ky_c} ~.  
\end{eqnarray}
We note that the exactly same factor as that in the gauge field solution 
Eq.~(\ref{xygauge}) comes out again in the gaugino solution.  

Of course, we can study every configuration 
considered already in the gauge boson cases  
for the brane fermionic kinetic terms and sources again. 
But here let us discuss only the case that a brane gaugino  
kinetic term, and a bulk and a brane sources are present.  
Then the equation of motion for the gaugino is      
\begin{eqnarray}
\bigg[\gamma_5k\hspace {-2.3mm}/-i\partial_y
+\delta(y)\frac{2y_cg^2}{e^2}\gamma_5k\hspace {-2.3mm}/\bigg]\tilde{\Psi}(k,y)
=2y_cg^2\gamma_5\bigg(\tilde{\eta}^B(k,y)+\tilde{\eta}^b(k)\delta(y)\bigg) ~.   
\end{eqnarray}
The solution has the form similar to the bosonic one 
Eq.~(\ref{bosonb})--(\ref{bosone}), 
\begin{eqnarray}
\Psi(x,y)&=&\Psi^B(x,y)+\Psi^b(x,y) ~, \\
\Psi^B(x,y)&=&\int \frac{d^4k}{(2\pi)^4}\sum_{k_5}e^{ikx}e^{ik_5y}
\frac{k\hspace{-2.3mm}/-k\hspace{-2.3mm}/_5}{k^2-k_5^2}
\cdot g^2\hat{\eta}^B(k,k_5) ~, \\
\Psi^b(x,y)&=&\int \frac{d^4k}{(2\pi)^4}\sum_{k_5}
e^{ikx}e^{ik_5y}\frac{k\hspace{-2.3mm}/-k\hspace{-2.3mm}/_5}{k^2-k_5^2}~
\frac{(e^2g^2 \tilde{\eta}^b(k)-g^4 \tilde{\zeta}^B(k))}
{e^2+g^2ky_c~{\rm cot}ky_c} ~, 
\end{eqnarray}
where $\hat{\eta}^B(k,k_5)$ and $\tilde{\zeta}^B(k)$ are defined as 
\begin{eqnarray}
\tilde{\eta}^B(k,y)&\equiv& \frac{1}{2y_c}\sum_{k_5}
\hat{\eta}^B(k,k_5)e^{ik_5y} ~,  \\
\tilde{\zeta}^B(k)&\equiv& \sum_{k_5}
\frac{k^2-k_5k\hspace{-2.3mm}/}{k^2-k_5^2}~\hat{\eta}^B(k,k_5) ~.  
\end{eqnarray}

%
%\begin{eqnarray}
%&&\frac{1}{g^2}\bigg(\partial_\mu^2-\partial_y^2
%+\frac{2y_cg^2}{e^2}\delta(y)\partial_\mu^2\bigg)G(x-x';y-y')
%=\delta^4(x-x')\delta(y-y') \\
%&&G(x-x';y-y')=-\int \frac{d^4k}{(2\pi)^4}\sum_{k_5}
%\frac{e^{ik(x-x')}e^{ik_5y}}{k^2-k_5^2}\bigg[g^2e^{-ik_5y'} 
%%\\
%%&&~~~~~
%-\frac{g^4\sum_{k_5'}\frac{k^2}{k^2-k_5^{'2}}e^{-ik_5'y'}}
%{e^2+g^2ky_c~{\rm cot}ky_c}
%\bigg] \nonumber \\
%\end{eqnarray}
%

\section{The model}

Now let us consider $S^1/Z_2$ orbifold compactificaton 
of the 5D, $SU(2)$ gauge theory.  
We assign $Z_2$ odd parity to charged generators $T^1$, $T^2$, and 
even parity to diagonal one $T^3$, which is consistent with the $SU(2)$ 
Lie algebra.   
Then $SU(2)$ gauge symmetry breaks down to $U(1)$, 
and only $U(1)$ is respected on the brane.  
In this setup, the low energy theory is just $U(1)$ gauge theory.  
On the other hand, above the compactification scale $\mu>1/y_c$, 
$SU(2)$ symmetry is restored approximately.  
%By another $Z_2'$, $N=2$ SUSY breaks to $N=1$.   

In addition ot this, we can introduce a gauge kinetic term 
on the $S^1/Z_2$ fixed point (brane) at $y=0$ having the form,  
\begin{eqnarray}
-\delta(y)\frac{1}{4e^2}F^3_{\mu\nu}F^{3\mu\nu} ~, 
\end{eqnarray} 
where $F^3_{\mu\nu}\equiv \partial_\mu A^3_\nu-\partial_\nu A^3_\mu$, 
and `3' is the group index of $SU(2)$.  
The additional $U(1)$ gauge kinetic term on the brane 
is not harmful in the model. 
Then, the solution of $A^3_\mu$ is given by  
Eq.~(\ref{bosonb})--(\ref{bosone}), while 
the solutions for $A^1_\mu$ and $A^2_\mu$ are the ordinary ones
in the absence of the brane kinetic term, Eq.~(\ref{bosonm}),     
since their wave functions vanish at the brane.
Additionally, we introduce some matter fields only on the brane.   

In this setup, we consider radiative corrections to the gauge boson masses 
and the gauge couplings.   
When we discuss radiative corrections in non-Abelian gauge theory,  
we should consider also the diagrams in which the ghost fields are involved 
in the loops.  To get the ghost Lagrangian, it is necessary to examine the 
gauge transformation.   
The bulk and brane gauge kinetic terms are invariant 
under the gauge transformation, 
\begin{eqnarray} \label{gaugetrf}
\delta A^a_M=\frac{1}{2y_cg^2}\bigg[
\partial_M\alpha^a-\epsilon^{abc}A^b_M\alpha^c
\bigg] 
+\frac{1}{e^2}\bigg[\partial_M\bigg(\alpha^a\delta^M_\mu\delta(y)\bigg)
-\epsilon^{abc}A^b_M\bigg(\alpha^c\delta^M_\mu\delta(y)\bigg)\bigg] ~, 
\end{eqnarray}
where $\alpha^a$ is the gauge parameter.   
Since odd (even) parity is assigned 
to $\alpha^{1}$, $\alpha^2$ ($\alpha^3$), 
$\alpha^1$, $\alpha^2$ vanish at $y=0$.  
Thus, non-vanishing term in the part proportional to $1/e^2$ 
in Eq. (\ref{gaugetrf}) is 
$\delta(y)\delta^M_\mu\partial_M\alpha^3/e^2$.  
Then, by the standard procedure, the ghost Lagrangian is obtained,    
\begin{eqnarray} \label{ghost}
{\cal L}_{\rm ghost}=-\frac{1}{2y_cg^2}~\bar{c}^a\partial^2_Mc^a
-\epsilon^{abc}\bar{c}^a\partial^MA^b_Mc^c
-\delta(y)\frac{1}{e^2}~\bar{c}^3\partial^2_\mu c^3 ~. 
\end{eqnarray}
Hence, the propagators of the ghost fields $c^1$, $c^2$, and 
$c^3$ are expected to have a form similar to    
those of $A^1_\mu$, $A^2_\mu$, and $A^3_\mu$, respectively.  

Let us consider one loop mass corrections 
to massless and massive gauge bosons.  
The mass corrections of KK modes $A^3_{n,\mu}$ ($n\neq 0$)  
result from the loops by  
massive $A^1_{n,\mu}$, $A^2_{n,\mu}$, $c_n^1$, and $c_n^2$ 
as seen in Fig. 2-(a), as well as matter fields living on the brane. 
Since odd parity is assigned to such gauge bosons and ghost fields, 
their propagators are the same exactly as those 
in the absence of the additional brane $U(1)$ gauge kinetic term.  
As shown in Ref. \cite{quiros}, all of quadratic divergences for the masses 
of $A^3_{n,\mu}$ cancel, but   
logarithmically divergent mass corrections to {\it massive} KK modes 
do not vanish, because the massive KK modes are not protected by 
gauge symmetry against such divergences.      
By the gauge symmetry, the massless mode $A_{0,\mu}$ remains massless 
\cite{quiros}.  
\begin{figure}
%[t]
\begin{center}
\includegraphics[width=70mm]{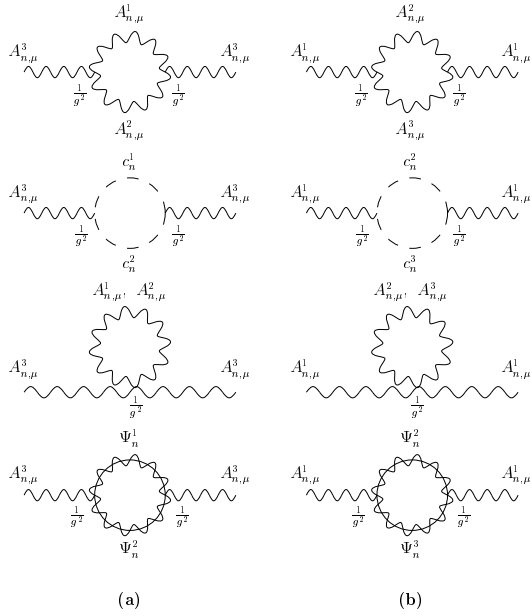}
\end{center}
\caption{
One loop corrections to the wave functions (or gauge couplings) 
and masses of ${\bf (a)}$ $A^3_{n,\mu}$, and ${\bf (b)}$ $A^1_{n,\mu}$.  
The third diagrams in ${\bf (a)}$ and ${\bf (b)}$ contribute only to 
the mass corrections of $A^3_{n,\mu}$ and $A^1_{n,\mu}$, respectively. }  
\end{figure}

The masses of $A^{1(2)}_{n,\mu}$ are corrected by 
$A^{2(1)}_{n,\mu}$, $A^3_{n,\mu}$, $c_n^{2(1)}$, and $c_n^3$ as in Fig. 2-(b).  
Since all of the self-interacting couplings of $SU(2)$ gauge fields 
are contained in the bulk kinetic term $F^a_{MN}F^{aMN}$, 
the propagator of $A^3_{n,\mu}$ should be given 
by Eq. (\ref{green}) rather than Eq. (\ref{green2}).  
Due to the second term in Eq. (\ref{green}), 
the loop corrections to masses of $A^1_{n,\mu}$ and $A^2_{n,\mu}$ 
are different from those of $A^3_{n,\mu}$.  
However, the second term in the propagator, Eq. (\ref{e^2}), 
does not give rise to quadratic divergences 
in the one loop mass corrections.   
Therefore, only if the cutoff scale is not very high above the compactification 
scale, could the KK modes masses be perturbatively stable.   

Next, let us examine RG running behavior of the gauge couplings.
The running behavior of $g_{eff}^2$ is different from $g^2$,            
because basically they are independent gauge couplings.
The radiative corrections to $g^2_{eff}$ arise   
from the first two diagrams in Fig. 2-(a) and the diagram by brane matter.  
Below the compactification scale, only $A^3_{0,\mu}$ is a massless 
gauge boson, and it only couples to matter.
However, above the compactification scale, KK modes,
$A^1_{n,\mu}$ and $A^2_{n,\mu}$ ($n\neq 0$) also take part in 
its correction \cite{dines}. 
This interaction could draw the effective gauge coupling
rapidly to much smaller value at higher energy.
However, we could imagine that linearly asymptotically free  
behavior of $g^2_{eff}$ above the compactification scale by KK gauge bosons 
could be mild by introducing many fields on the brane.  

On the other hand, the propagations of $A^1_{n,\mu}$ and $A^2_{n,\mu}$ 
are associated only with $g^2$.  Above the compactification scale, 
$g^2$ also would falls down rapidly to smaller value at higher energy scale 
by the gauge bosons and the ghost fields 
contributions to the loops in Fig. 2-(b). 
%
%With Eq.~(\ref{bosonm}), (\ref{}) and (\ref{ghost}) 
%one can calculate the diagrams in Fig. 2.--(a) and (b),
%\begin{eqnarray} \label{loop3}
%\Pi^{3,3}_{\mu\nu}&=&-\frac{1}{8\pi^2\epsilon}
%\bigg(k_\mu k_\nu -g_{\mu\nu}k^2\bigg)\cdot \frac{10}{3} ~, \\
%\Pi^{i,j}_{\mu\nu}&=&-\frac{\delta^{ij}}{8\pi^2\epsilon}
%\bigg(k_\mu k_\nu -g_{\mu\nu}k^2\bigg)\cdot \frac{10}{3}
%~\frac{g^2_{n,eff}}{g^2} ~, \label{loop12} 
%\end{eqnarray}
%where $i, j=1,2$.  
%

We can generalize our scenario to a supersymmetric system by introducing 
superpartners in the bulk and on the brane.  
We introduce another $Z_2'$ to break $N=2$ SUSY into $N=1$.  
As in the case of the gauge bosons, the propagation of the gauginos 
$\Psi_n^1$ and $\Psi_n^2$ still respect $g^2$, 
while that of $\Psi_0^3$ does $g_{eff}^2$ at low energy.  
If supersymmetric chiral matter fields are assumed to be only on the brane, 
the blow-up of the $g_{eff}^2$ at low energy could be mild, but they 
would not affect $g^2$. 
Actually, the smallness of $g_{eff}^2$ is protected by $e^2$.  
Thus, possibly, $g_{eff}^2$ remains small at low energy, while $g^2$ blows up 
above the compactication scale, $1/y_c<\mu<\Lambda_{\rm cutoff}$.  
Then, $\Psi^1$ and $\Psi^2$ could condense, but 
$\Psi^3$ still composes an $N=1$ supermutiplet together with $A^3_\mu$.   
Hence, $U(1)$ gauge symmetry and $N=1$ supersymmetry are expected 
to survive at low energy. 
However, the condensations of $\Psi_n^1$ and $\Psi_n^2$ lead to  
breaking of remainig $N=1$ SUSY softly by gravity mediation. 
For example, in the off-shell formalism of 5D supergravity \cite{zucker}, 
\begin{eqnarray}
{\cal L}_{SUGRA}&=&\frac{1}{2y_c}\bigg[\cdots -12\vec{t}^2~+\cdots
-\kappa\bar{\Psi}^a\vec{\tau}\Psi^a~\vec{t}~+ \cdots \bigg] ~,  
%\\&&~~+\delta(y)\bigg[\cdots-48\kappa_*^2(w-1)~|\phi|^2~\bigg((t^1)^2
%+(t^2)^2\bigg)\cdots \bigg]~, \nonumber
\end{eqnarray}
in which $\kappa$  
%
%and $\kappa_*$ are 
%
is defined as the Planck   
%
% and the fundamental 
%
scale mass parameters, 
$\kappa\equiv 1/M_P$.  
%
%, and 
%
%$\kappa_*\equiv 1/M_*$, respectively.  
%
$\vec{\tau}$ indicates $SU(2)_R$ generators.  
%
%, and $w$ denotes the $R$ charge 
%of the scalar field $\phi$, which belongs to the chiral multiplet localized 
%on the brane.   
%
$\vec{t}$ is a triplet auxiliary field in the gravity multiplet.  
After eliminating the auxiliary fields by using their equation of motion, 
it is shown explicitly that the mass term of $\Psi^3$ 
is generated only if $\langle\Psi^1\vec{\tau}\Psi^1\rangle 
=\langle\Psi^2\vec{\tau}\Psi^2\rangle\neq 0$.      
SUSY breaking of the gauge sector leads to SUSY breaking of the chiral 
multiplet on the brane \cite{antoniadis}.   

Hence, in this scenario, we don't have to introduce 
an additional hidden sector.   
SUSY of the visible sector could be broken softly 
without an additional hidden sector, 
because the fields with odd parity play the role of 
the confining hidden sector in the gravity mediation SUSY breaking scenario.  
This scenario works because basically 
the interactions by the odd parity gauge fields, and by the even parity 
gauge fields are different, due to the additional brane kinetic term.  
We could generalize this simple model to theories with larger gauge 
symmetries.   
 
%We could also imagine another interesting scenario.  
%Instead of $SU(2)$, let us assmue the orbifold braeking of larger 
%gauge gourp such that remaining gauge group is non-Abelian. 
%Then we can make two different gaugino confining scales, 
%$\langle\lambda^{\rm odd}\lambda^{\rm odd}\rangle$ 
%and $\langle\lambda^{\rm even}\lambda^{\rm even}\rangle$.  
%Then, we can introduce two independent supersymmetry breaking scale 
%at low energy \cite{kyae}.    

\section{Conclusion}

In conclusion, we obtain various particular solutions of the gauge field and 
gaugino in the situation where there are additional brane kinetic terms 
as well as 
the bulk gauge kinetic terms.  While the gauge interaction at low energy is 
universal, above the compactification scale 
the gauge interactions on the brane and in the bulk are different.  
Since $N=2$ supersymmetric $SU(2)$ gauge theory 
reduces to an $N=1$ supersymmetric $U(1)$ gauge theory at low energy 
by $S^1/(Z_2\times Z_2')$ orbifolding, 
we could introduce 
additional brane kinetic terms respecting $N=1$ SUSY and 
$U(1)$ gauge symmetry at a $Z_2$ fixed point.   
In such a model,  
even if the bulk gauge coupling $g^2$, which is asymptotically free, 
blew up above comapactification scale,  
the low energy effective $U(1)$ gauge coupling $e^2g^2/(e^2+g^2)$ 
could remain small,   
because they are basically independent parameters.   
Since the heavy gauginos with odd parity under $Z_2$ 
interact with $g^2$, the blowing up of $g^2$ could make them condensed.    
By gravity mediation, the effect   
triggers the soft mass terms of the gaugino with even parity.

\vskip 0.3cm

\noindent {\bf Acknowledgments}

\noindent 
The author thanks S. M. Barr, Q. Shafi, J. E. Kim, and Kyuwan Hwang 
for reading the manuscript and discussions.
The work is partially supported
by DOE under contract number DE-FG02-91ER40626.

\end{document}